# Acoustic Signal Analysis with Deep Neural Network for Detecting Fault Diagnosis in Industrial Machines

M.YURDAKUL, Ş. TAŞDEMİR

Kırıkkale University, Kırıkkale TURKEY

Selçuk University, Konya TURKEY

myyurdakul@gmail.com

stasdemir@selcuk.edu.tr

Abstract: Detecting machine malfunctions at an early stage is crucial for reducing interruptions in operational processes within industrial settings. Recently, the deep learning approach has started to be preferred for the detection of failures in machines. Deep learning provides an effective solution in fault detection processes thanks to automatic feature extraction. In this study, a deep learning-based system was designed to analyze the sound signals produced by industrial machines. Acoustic sound signals were converted into Mel spectrograms. For the purpose of classifying spectrogram images, the DenseNet-169 model, a deep learning architecture recognized for its effectiveness in image classification tasks, was used. The model was trained using the transfer learning method on the MIMII dataset including sounds from four types of industrial machines. The results showed that the proposed method reached an accuracy rate varying between 97.17% and 99.87% at different Sound Noise Rate levels.

Keywords; early fault diagnosis, industrial machinery, convolutional neural networks, deep learning, MIMII dataset, transfer learning, fine-tuning.

1. INTRODUCTION

Advancements in industrial technology have enabled factories to become more efficient and intelligent, thereby transforming them into systems that operate continuously, 24 hours a day [1-3]. However, continuous operation can lead to mechanical failures in machines, cause financial losses, and even jeopardize human safety [3, 4]. Deep Learning(DL) is one of the most popular and effective machine learning techniques of recent years. DL is used to analyze various types of data, including sound and image [5-7]. The analysis of sounds produced by machines during production in factories can aid in the early detection of failures [8]. By analyzing sounds, deep learning can enable early detection of failures, reducing costs, increasing safety, and eliminating the need to manually monitor machines. As a result, it reduces human workload and increases work efficiency. It provides more accurate data for factories and can improve their operational performance and efficiency. MIMII (Malfunctioning Industrial Machine Investigation and Inspection) [9] is a dataset created for identifying the sounds of faulty machines in a factory, and it can be used as a benchmark for the detection of sound-based machine malfunctions. In this study, a method was presented using the MIMII dataset to identify normal and abnormal states of machines in a factory.

The main contributions of this study can be summarized as follows:

- • A method was proposed to classify sound data obtained from machines in a factory environment into normal and abnormal categories. First, Mel spectrograms were created from the sound data and an image dataset based on these spectrograms was established.
- • The DenseNet169 Convolutional Neural Network (CNN) model was chosen to separate Mel spectrogram images into normal and abnormal categories.
- • The classification performance at different Sound Noise Rates was compared.

The proposed method is a promising approach to detect abnormal conditions in factory machines using sound data. This method has the potential to offer significant benefits such as increased productivity, reduced maintenance costs, and the creation of a more precise and safe working environment.

The rest of the article is organized as follows: In Section 2, existing studies in the literature are reviewed. In Section 3, the methods used, including the dataset and the proposed classification system, are explained. Section 4 presents and analyzes the results of the study. Finally, in Section 5, the article is concluded and suggestions for future research are provided.

## 2. RELATED WORKS

In recent years, there has been increasing interest in the use of acoustic signals in diagnosing machine failures, as these signals provide valuable information about a machine's condition. Numerous studies have been conducted for the analysis of acoustic signals coming from machines, using various techniques ranging from traditional signal processing methods to deep learning approaches.

Henze et al. (2019) proposed the AudioForesight method, a sound-based approach for the detection of industrial machine failures[10]. The AudioForesight system uses a combination of anomaly detection and classification techniques to monitor and identify system failures based on pre-defined error classes.

Becker et al. (2020) proposed a machine learning system for detecting flaws and errors in a Fused Deposition Modeling printer using acoustic signals[11]. In the study, audio samples are recorded and labeled, data augmentation techniques are applied, and features are extracted using Mel-frequency cepstral coefficients. Mel-frequency cepstral coefficients were utilized in training a Long Short-Term Memory (LSTM) model on various classes of relevant sounds that occur during 3D printing.

Duman et al. (2020) introduced a novel unsupervised Acoustic Anomaly Detection(AAD) system that based on a Convolutional Auto Encoder (CAE) to detect abnormalities in industrial processes from acoustic data [12]. The results showed that CAE outperformed One-Class Support Vector Machine (OCSVM), and the hybrid approach performed similarly to CAE.

Koizumi et al. (2020) proposed SPIDERnet, a deep neural network for detecting faults in industrial machinery using acoustic signals[13]. The study shows that SPIDERnet outperforms conventional methods and provides insights into the use of deep neural networks for fault detection. This work can inspire future research in the field of fault diagnosis and prognosis in industrial settings.

Gantert et al. (2021) compared the performance of two methods, SVM and MLP, for detecting faults in industrial equipment using acoustic signals [14]. The study utilized statistical features extracted from the signals as input to the classifiers. The results showed that MLP outperformed SVM in detecting faults, indicating the effectiveness of the proposed method for fault detection.

Tama et al. (2022) proposed an ensemble CNN method for diagnosing valve and pump faults using sound signals[15]. The method consisted of combining multiple CNNs trained on different subsets of the dataset to improve the accuracy of fault detection. The study shows that the proposed ensemble model outperforms other models for detecting malfunctions in pumps and valves.

Lyu et al. (2022) introduced a novel deep learning method, RSG, for accurately diagnosing bearing faults in high-noise industrial environments[16]. The RSG approach integrates residual building units, soft thresholding, and global context mechanisms to extract informative features from vibration signals. Experimental results showed the effectiveness of RSG, attaining an average diagnosis accuracy of 98% on a dataset of faulty motors.

3. MATERIAL AND METHODS

Industrial machines emit sounds while operating and the analysis of these sounds provides valuable information about the condition of the machine. In this study, a method is proposed for the classification of normal and abnormal sounds in industrial machines based on acoustic signals. The MIMII dataset, which contains sounds of 4 different machines under normal and abnormal conditions, was used in the study. Feature extraction is a crucial step that can significantly affect the accuracy of the classification process. Mel spectrogram is a type of spectrogram that visually represents the frequency components of sound signals over time.

Many studies[17-22] have shown that using mel spectrograms as a feature extraction method in audio analysis can be very effective, and can lead to successful outcomes in different applications. Therefore, in this study, mel spectrograms were used to extract features from sound data in the MIMII dataset and an image dataset is created using these spectrograms.

DenseNet169 [23] architecture is used to classify machine sounds into normal and abnormal conditions. The model is trained using transfer learning and fine-tuning techniques. The diagram of the proposed study is shown in Figure 1.

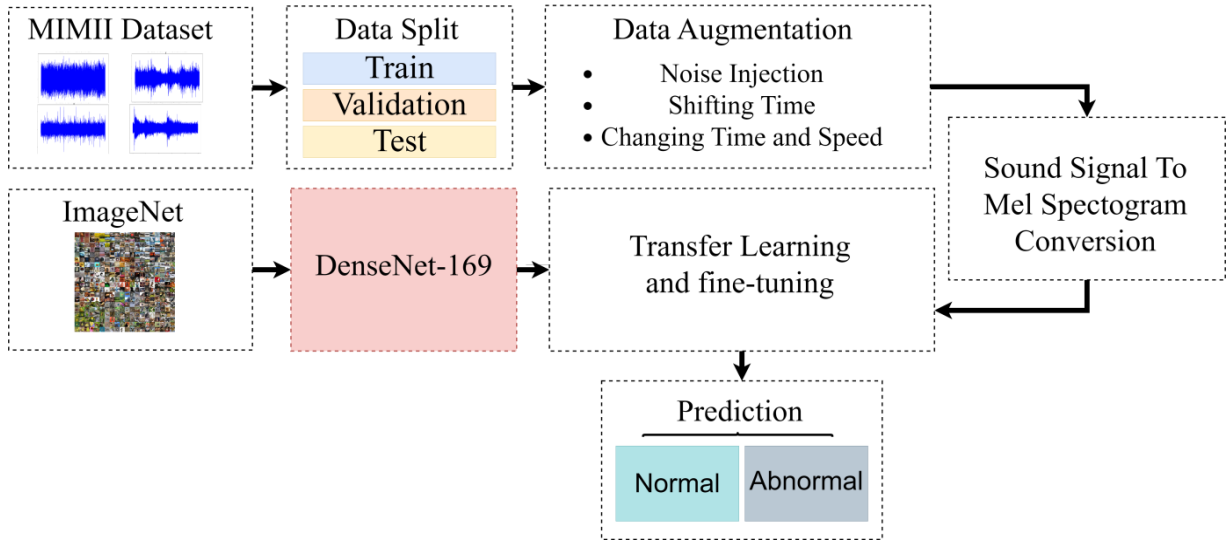

**Fig. 1.** Schematic diagram of proposed method

### 3.1. Dataset

The MIMII (Malfunctioning Industrial Machine Investigation and Inspection)[9] dataset is a benchmark dataset used in sound-based fault detection, which includes normal and abnormal sounds from machines in a factory environment. The dataset contains sounds from four different machines: sliders, pumps, valves, and fans. The valves are solenoid valves that continuously opening and closing, while the pump is used for water intake and discharge. The fans are industrial fans used for gas or air flow. The slider is a system with continuous linear motion. The sounds were recorded in 16-bit format with an 8-channel microphone structure at a sampling rate of 16 kHz. The microphone setup was placed 50 cm away from the pump, fan, and slider, and 10 cm away from the valve, and a 10-second audio recording was captured.

Additionally, the dataset contains data for three different Signal Noise Rates (SNR): -6, 0, and 6 dB. The number of data is equal for all SNR conditions. The data for each SNR level is presented in Table 1. Figures 3, 4, and 5 shows sample images that distinct levels of Signal-to-Noise Ratio (SNR) conditions in the dataset. The data was divided into three parts, with 80% for training, 10% for validation, and 10% for testing. Afterwards, data augmentation techniques were employed to balance the normal and abnormal data.

**Table 1.** MIMII Dataset Content Details of Different SNR conditions

|  | Normal | Abnormal |
|---|---|---|
| **Fan** | 4075 | 1475 |
| **Slider** | 3204 | 890 |
| **Pump** | 3749 | 456 |
| **Valve** | 3691 | 479 |

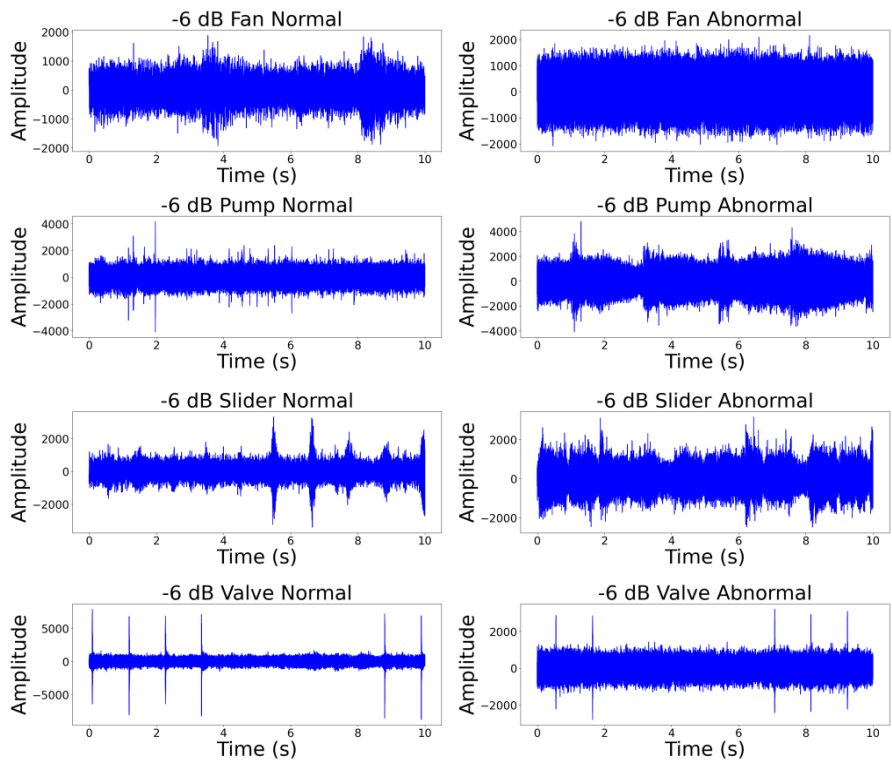

**Fig. 3.** -6 dB Machine Sound Waveform

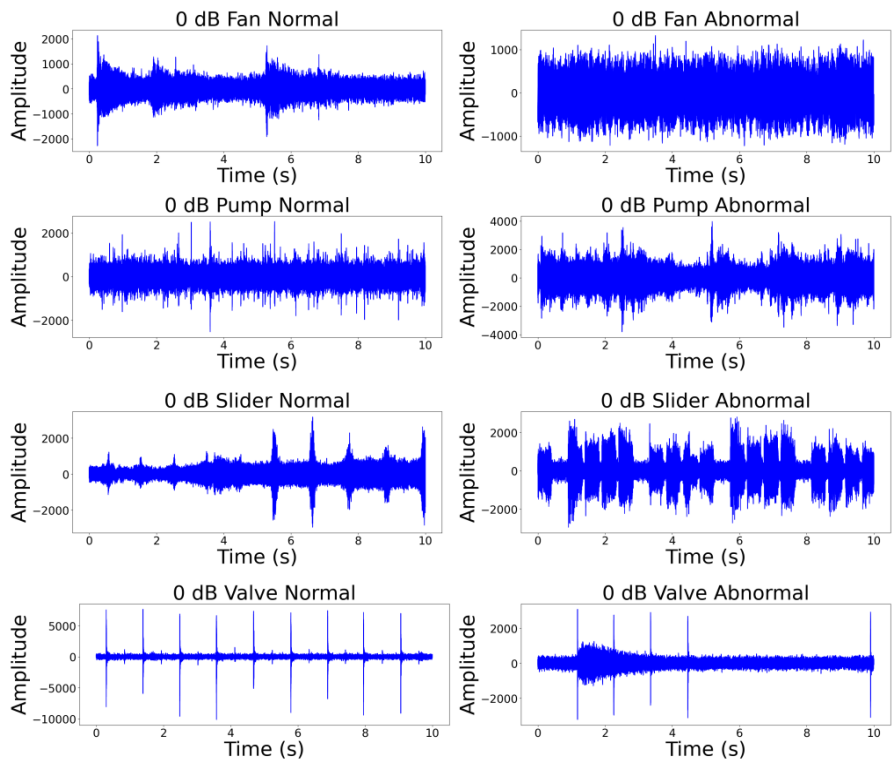

**Fig. 4.** 0 dB Machine Sound Waveform

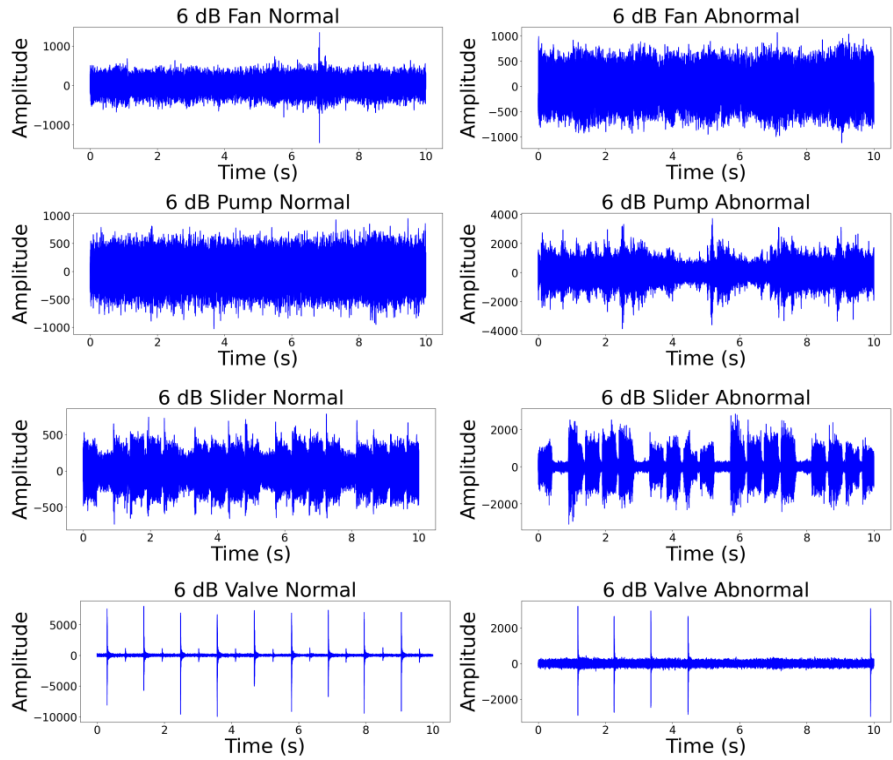

**Fig. 5.** 6 dB Machine Sound Waveform

### 3.2. Data Augmentation

In the MIMII dataset, there is an imbalance in the distribution of normal and abnormal sound data. In order to balance data, the amount of data for abnormal sound data has been increased by using Noise Injection, Shifting Time, Changing Pitch, and Changing Speed methods.

• Noise Injection, random noise is added to audio data through the injection method. The data with added noise can improve the network better recognize anomalies or rare cases.

• Shifting time is a simple audio augmentation technique that shifts an audio randomly to the right or left by a certain amount of time.

• Changing pitch method aims to change the length and tone of audio data by altering its Pitch. Changing pitch involves altering the fundamental frequency of audio, which shifts its overall pitch

• Changing speed method aims to change the speed of audio data. This method can be used to increase or decrease the speed of audio data.

The distribution of data after the data augmentation process can be viewed in Table 2.

Table 2. MIMII dataset details with augmented data

|  | Normal | Anormal |
|---|---|---|
| **Fan** | 4075 | 4039 |
| **Slider** | 3204 | 3179 |
| **Pump** | 3749 | 3747 |
| **Valve** | 3691 | 3683 |

### 3.3. Feature Extraction

The main goal of this study is to utilize convolutional neural networks (CNN) for the classification of machine sounds as either normal or abnormal. CNNs are specifically designed to process and analyze image data. Therefore, in order to use CNNs for sound classification, the sound data must be transformed into a visual format. In this context, mel Spectrogram method was used. Mel spectrogram uses the mel scale to measure the intensity of different frequencies in a sound wave and creates a visual representation of the frequencies over time called a spectrogram. After the data augmentation, Mel Spectrograms were generated for each sound data. Figure 6 displays sample Mel Spectrograms of the dataset, which illustrate the differences in visual patterns for varying Signal-to-Noise Ratios (SNRs).

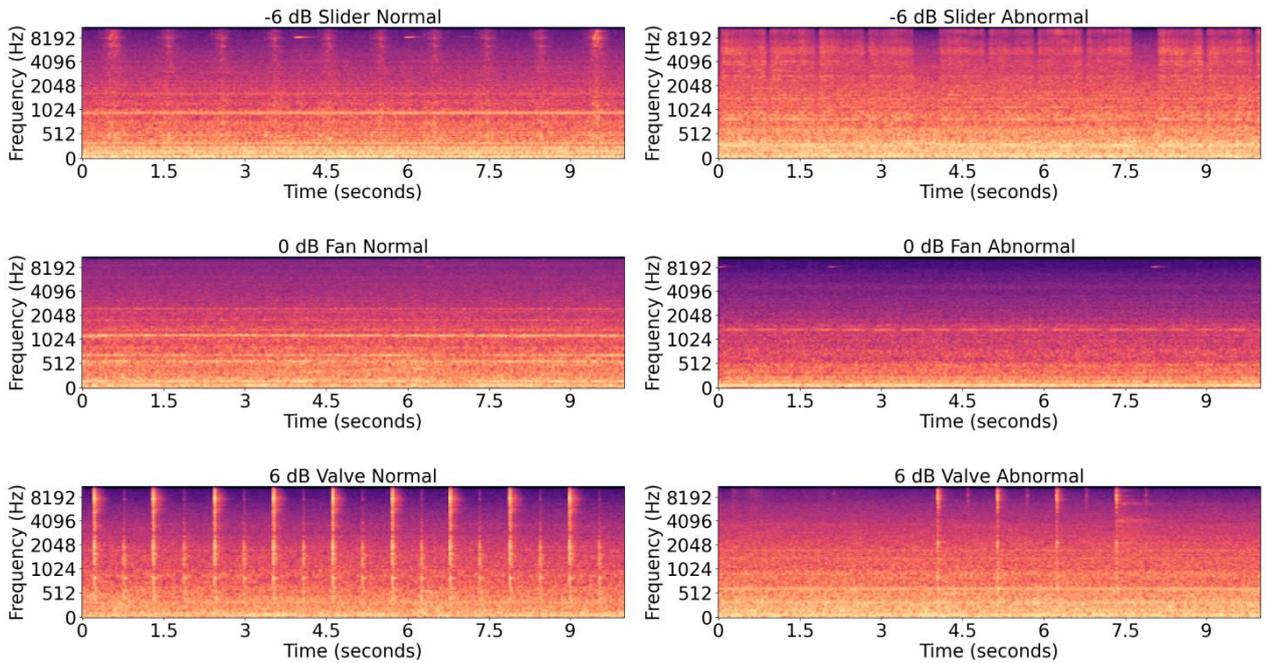

**Fig. 6.** Mel-Frequency Spectrograms of Sample Audio Signals

### 3.4. Convolutional Neural Network Model

Convolutional Neural Network is a successful deep learning algorithm widely used in image-based analysis studies. CNNs have been applied in studies with impressive results in various domains such as medicine[24], education[25], manufacturing[26], music[27], defense[28], and agriculture[29]. In contrast to classical image analysis approaches, CNNs have additional layers that can perform automatic feature extraction from images. CNNs consists of three main components: convolutional, pooling, and fully connected. In the convolutional layer, important features representing the image are extracted by applying kernels to the input layers.

The pooling layer is used to increase computational efficiency and learning speed by reducing the size of the features produced by the convolutional layer.

In the fully connected layer, the input is a feature vector, which is fed into the artificial neural network to make prediction.

Numerous architectures have been proposed to enhance the accuracy of CNNs by utilizing diverse combinations and quantities of the fundamental layers.

These architectures aim to optimize the feature extraction and classification capabilities of CNNs. In the literature, there are many popular CNN models such as LeNet, AlexNet, ZfNet, VGG, GoogleNet, ResNet, DenseNet and Xception, [30]. DenseNet[23] algorithms proposed by G. Huang and his team. This model was accepted as the winner of the ImageNet Competition in 2017. DenseNet allows for additional inputs collected from previous layers to be transferred to all subsequent layers. Each layer collects information from higher levels. Due to the fact that the network receives features from all previous layers, it may be narrow and tighter, resulting in a decrease in the number of channels (and thus an improvement in calculation and memory efficiency). DenseNet-169 is a member of DenseNet algoritms which has 169 layers. In this study, transfer learning is used with the DenseNet-169 algorithm to classify normal and abnormal sounds. At the end of the model, a Global Average Pooling layer was added, followed by a classification layer using the sigmoid function. The model was trained for 25 epochs using the SGD optimizer and binary cross entropy loss function.

### 3.5. Performance Metrics

Performance evaluation metrics for CNN models can be calculated using the confusion matrix. Metrics include accuracy, precision, recall, F1-score, Kappa, MCC, and AUC. Accuracy is the ratio of correct predictions to total predictions. Precision is the ratio of true positives to all positive predictions, and recall is the ratio of true positives to actual positive cases. The F1-score is the harmonic mean of precision and recall. Kappa measures agreement between predicted and actual outcomes. AUC represents the relationship between positive and negative labels in the dataset and their prediction by the model. MCC measures the correlation between predicted and actual outcomes. These metrics provide an overall understanding of model performance, and can be used to compare different models and identify strengths and weaknesses. TP, TN, FP, and FN are concepts used in measuring the performance of a classification model, where TP indicates the number of positive samples correctly classified as positive, TN indicates the number of negative samples correctly classified as negative, FP indicates the number of negative samples incorrectly classified as positive, and FN indicates the number of positive samples incorrectly classified as negative. Equations of metrics can be seen between Equations 1-6.

$$Accuracy = \frac{TP+TN}{TP+FN+FP+TN} \qquad (1)$$

$$Precision = \frac{TP}{TP+FP} \tag{2}$$

$$Recall = \frac{TP}{TP+FN} \tag{3}$$

$$F1 - Score = \frac{2*Precision*Recall}{Precision+Recall} \tag{4}$$

$$Kappa = \frac{2*(TP*TN-FN*FP)}{(TP+FP)*(FP+TN)*(TP+FN)*(TN+FN)} \tag{5}$$

$$MCC = \frac{TP*TN-FP*FN}{\sqrt{(TP+FN)*(TP+FP)*(TN+FP)*(TN+FN)}} \tag{6}$$

## 4. RESULTS AND DISCUSSION

### 4.1. Experiment Environment

The experiments were conducted using the Google Colab platform with 13 GB of RAM and the Keras 2.9.0 API with Tensorflow 2.9.2 as the backend, Python 3.8.10, Nvidia Tesla T4 GPU, and CUDA V11.2 for GPU acceleration. The results were visualized using the matplotlib 3.2.2 library.

### 4.2. Results

As a result, in this study, a method was proposed for the analysis of acoustic signals from industrial machines. A CNN model was trained using the MIMII dataset to classify machine sounds in normal and abnormal conditions. The DenseNet169 model was trained using transfer learning and fine-tuning techniques, achieved a ranging from %97.10 to 99.87% accuracy rate at various SNR levels. All machine classes achieved a perfect accuracy rate of 99.87% at the highest SNR level of 6 dB. However, at the lowest SNR level of -6 dB, the fan machine had the lowest accuracy rate among all the machine classes tested. These results demonstrate the effectiveness of deep learning in the analysis of acoustic signals from industrial machines. The proposed method can minimize operational disruptions and increase productivity by enabling early detection of machine malfunctions. Additionally, it can reduce the need for manual listening and observation, thus reducing workload and increasing work efficiency. Table 3 shows the results of the classification, which was conducted on four different industrial machines at three different Signal-to-Noise Ratio (SNR) levels (-6dB, 0dB, and 6dB).

**Table 3.** Evaluation metric results for different machines at different SNR level

| SNR | Machine | Accuracy | Precision | Recall | F1 Score | Kappa | MCC | AUC |
|---|---|---|---|---|---|---|---|---|
| -6 dB | Fan | 0.97170 | 0.97171 | 0.97171 | 0.97171 | 0.94341 | 0.94342 | 0.97170 |
| | Pump | 0.99202 | 0.99215 | 0.99202 | 0.99202 | 0.98404 | 0.98416 | 0.99202 |
| | Slider | 0.99687 | 0.99687 | 0.99687 | 0.99687 | 0.99374 | 0.99374 | 0.99687 |
| | Valve | 0.99864 | 0.99865 | 0.99865 | 0.99865 | 0.99729 | 0.99729 | 0.99864 |
| 0 dB | Fan | 0.99876 | 0.99877 | 0.99877 | 0.99877 | 0.99753 | 0.99754 | 0.99877 |
| | Pump | 0.99867 | 0.99867 | 0.99867 | 0.99867 | 0.99734 | 0.99734 | 0.99867 |
| | Slider | 0.99844 | 0.99844 | 0.99844 | 0.99844 | 0.99687 | 0.99687 | 0.99844 |
| | Valve | 0.99729 | 0.99731 | 0.99729 | 0.99729 | 0.99458 | 0.99460 | 0.99728 |
| 6 dB | Fan | **0.99877** | **0.99877** | **0.99877** | **0.99877** | **0.99753** | **0.99754** | **0.99876** |
| | Pump | **0.99867** | **0.99867** | **0.99867** | **0.99867** | **0.99734** | **0.99734** | **0.99867** |

| | | | | | | | |
|---|---|---|---|---|---|---|---|
| Slider | 0.99844 | 0.99844 | 0.99844 | 0.99844 | 0.99687 | 0.99687 | 0.99843 |
| Valve | 0.99865 | 0.99865 | 0.99865 | 0.99865 | 0.99729 | 0.99729 | 0.99864 |

The accuracy and loss metrics of the proposed model during the training process are illustrated in Figures 7, 8, and 9, encompassing various industrial machines and SNR levels.

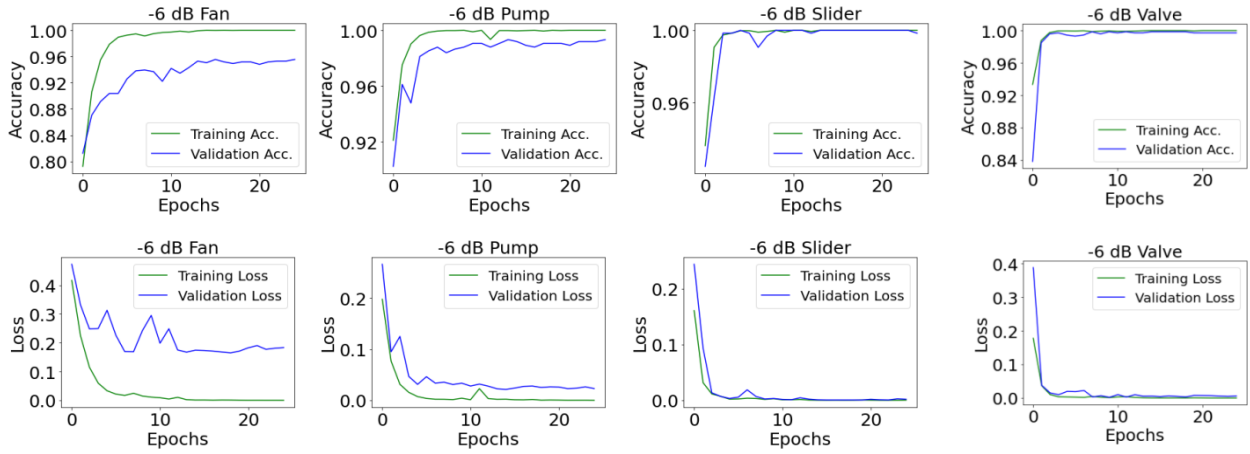

**Fig. 7.** -6 dB SNR Accuracy and loss graphics

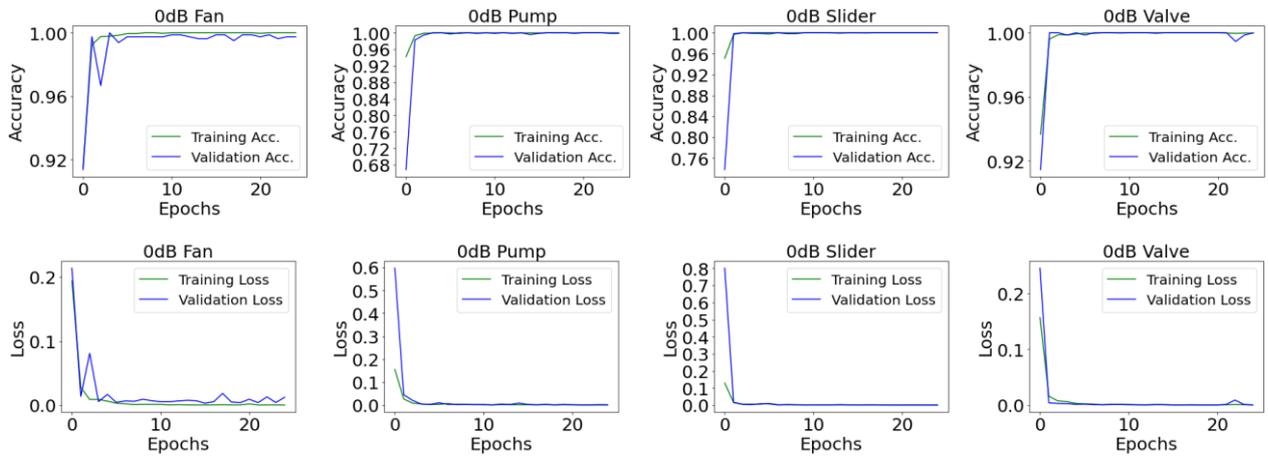

**Fig. 8.** 0 dB SNR Accuracy and loss graphics

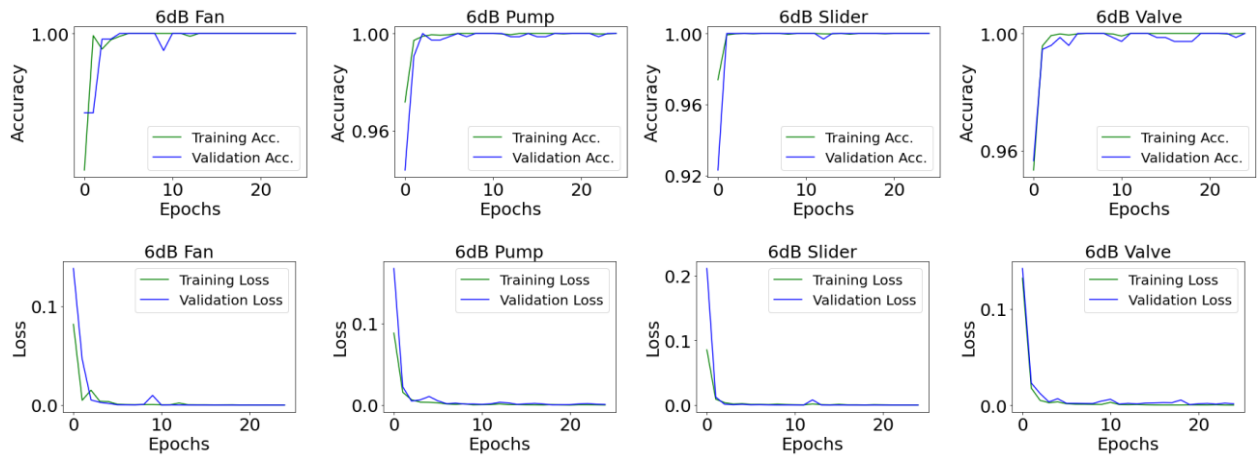

**Fig. 9.** 6 dB SNR Accuracy and loss graphics

### 4.3. Discussion

The confusion matrices presented in Figure 10, obtained at different SNRs, provide a detailed analysis of the proposed method's performance in classifying normal and abnormal sounds. The results indicate that the proposed method performs better in less noisy environments, as reflected by higher accuracy, precision, recall, and F1-score values at higher SNRs. However, the accuracy decreases and false positives increase at lower SNRs, making it hard to distinguish between normal and abnormal sounds due to the more prominent background noise. In order to improve the proposed method's performance, additional feature extraction techniques and pre-processing steps can be used, such as denoising filters, feature normalization, and time-frequency analysis. These techniques can help reduce the impact of background noise and enhance the accuracy of classification. Furthermore, the proposed method can be extended to other types of industrial machinery to facilitate early fault detection and prevent operational downtime.

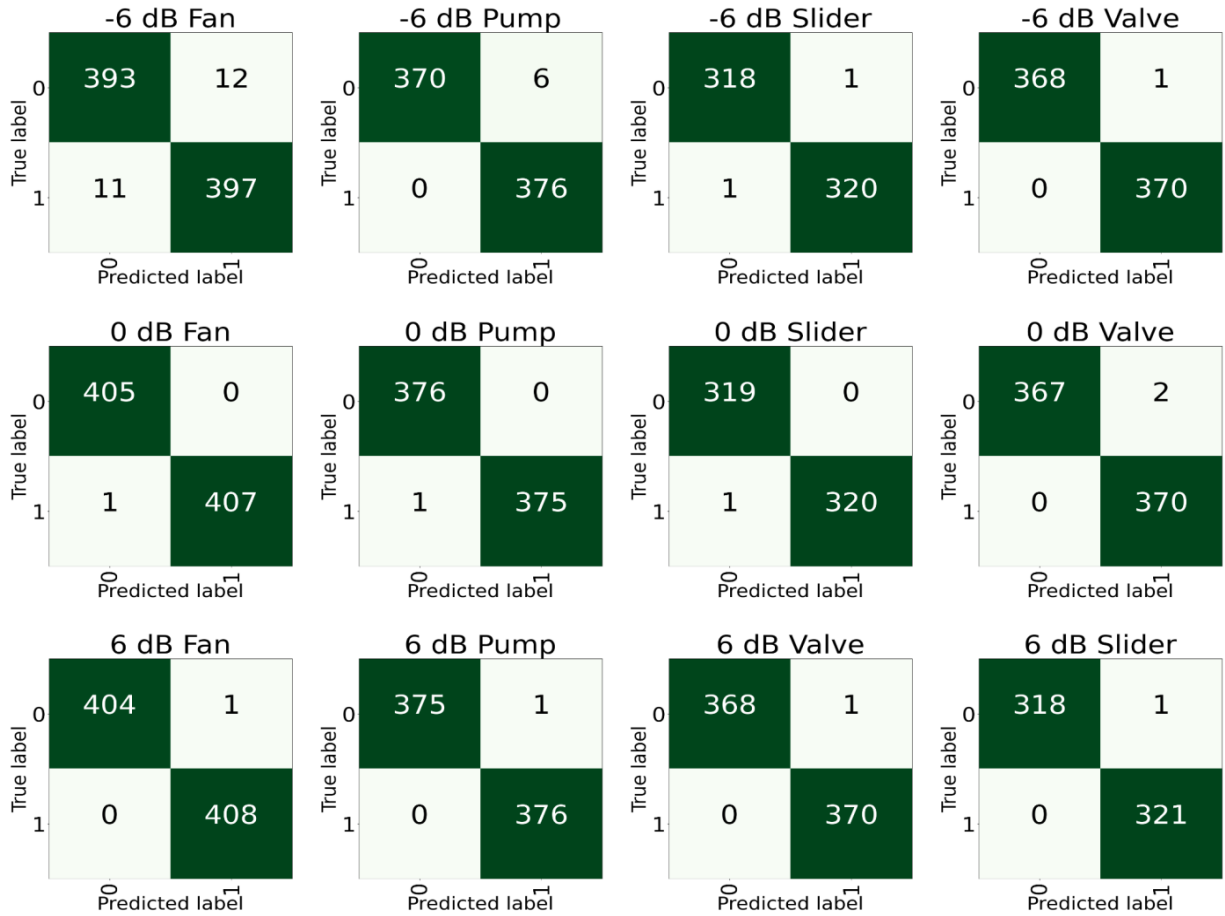

**Fig. 10.** Confusion Matrices of Models

In summary, the proposed method demonstrates the effectiveness of deep learning in analyzing acoustic signals from industrial machinery. It has the potential to minimize operational downtime, enhance productivity, and reduce the workload of manual inspection. Table 4-5 shows a comprehensive comparison of the proposed method's performance with existing studies for the classification of industrial machine sounds. Table 4 includes a variety of performance metrics, such as accuracy, precision, recall, and F1-score, for each method at different SNRs. In Table 5, only the AUC value is presented as it is a commonly used evaluation metric in existing studies. The results in Table 4 and 5 demonstrate that the proposed method outperforms many of the existing methods in terms of accuracy and other performance metrics, particularly at higher SNRs.

**Table 4.** Performance of existing studies in terms of various metrics

| Study | Technique | SNR | Machine | Accuracy | Precision | Recall | F1 Score | Kappa | MCC | AUC |
| --- | --- | --- | --- | --- | --- | --- | --- | --- | --- | --- |
| [15] | Ensemble CNN | - | Valve | 0.98211 | 0.98221 | 0.98357 | 0.98289 | 0.96414 | 0.96414 | 0.99930 |
|  |  | - | Pump | 0.97447 | 0.98198 | 0.96935 | 0.97562 | 0.94884 | 0.94893 | 0.99810 |

Table 5. Performance of existing studies in terms of AUC metric

| Study | Technique | SNR | Machine | AUC |
|---|---|---|---|---|
| [14] | MLP | 6 | Fan | 0.997 |
| | | 0 | | 0.977 |
| | | -6 | | 0.917 |
| | | 6 | Pump | 0.983 |
| | | 0 | | 0.966 |
| | | -6 | | 0.928 |
| | | 6 | Slider | 0.994 |
| | | 0 | | 0.985 |
| | | -6 | | 0.961 |
| | | 6 | Valve | 0.929 |
| | | 0 | | 0.842 |
| | | -6 | | 0.766 |
| | SVM | 6 | Fan | 0.991 |
| | | 0 | | 0.952 |
| | | -6 | | 0.86 |
| | | 6 | Pump | 0.936 |
| | | 0 | | 0.912 |
| | | -6 | | 0.849 |
| | | 6 | Slider | 0.971 |
| | | 0 | | 0.948 |
| | | -6 | | 0.901 |
| | | 6 | Valve | 0.802 |
| | | 0 | | 0.733 |
| | | -6 | | 0.677 |
| [13] | SPIDERnet | 0 | Fan | 0.994 |
| | | | Pump | 0.952 |
| | | | Slider | 0.979 |
| [9] | Auto Encoder | 6 | Fan | 0.94 |
| | | 0 | | 0.84 |
| | | -6 | | 0.70 |
| | | 6 | Pump | 0.81 |
| | | 0 | | 0.74 |
| | | -6 | | 0.68 |
| | | 6 | Slider | 0.90 |
| | | 0 | | 0.80 |
| | | -6 | | 0.70 |
| | | 0 | Valve | 0.67 |
| | | 6 | | 0.61 |
| | | -6 | | 0.53 |

## 5. Conclusion and Future Works

This study has demonstrated promising outcomes in analyzing acoustic signals from industrial machines with DenseNet-169 model. This approach could be enhanced in the future by utilizing more extensive datasets to improve the model's precision and robustness. Furthermore, the current approach can be adapted to analyze other industrial machines, such as turbines, compressors, and generators. Moreover, the proposed technique can be further developed to estimate the remaining useful life of industrial machines and predict their failure time. This can be accomplished by integrating time series analysis and recurrent neural networks to model the machine's degradation over time. Furthermore, merging the proposed acoustic signal analysis with other sensor data, such as temperature, vibration, and current, could provide a more comprehensive and precise representation of the machine's health status. In conclusion, the proposed approach has the potential to enhance the efficiency and productivity of industrial systems by identifying faults early and reducing operational downtime. Furthermore, future research in this area could contribute to developing more efficient and intelligent maintenance systems, advancing Industry 4.0.